\documentclass[prl,preprint,superscriptaddress,showpacs,aps]{revtex4-1}
\usepackage{amsfonts,amsmath,amssymb,bm}
\usepackage{graphicx,epsfig}
\usepackage{hyperref}
\usepackage{color}
\usepackage{colortbl}
\usepackage{hyphenat}
\usepackage{ulem}

\begin{document}

\title{Two-color coherent control of femtosecond above-threshold photoemission from a tungsten nanotip}

\author{Michael F\"orster}\email{michael.foerster@fau.de}
\affiliation{Department of Physics, Friedrich-Alexander-Universit\"at Erlangen-N\"urnberg (FAU), Staudtstra{\ss}e 1, 91058 Erlangen, Germany, EU}
\affiliation{Max Planck Institute of Quantum Optics, Hans-Kopfermann-Stra{\ss}e 1, 85748 Garching, Germany, EU}
\author{Timo Paschen}
\affiliation{Department of Physics, Friedrich-Alexander-Universit\"at Erlangen-N\"urnberg (FAU), Staudtstra{\ss}e 1, 91058 Erlangen, Germany, EU}
\author{Michael Kr\"uger}
\altaffiliation{Present address: Weizmann Institute of Science, 234 Herzl St., Rehovot 7610001, Israel}
\affiliation{Department of Physics, Friedrich-Alexander-Universit\"at Erlangen-N\"urnberg (FAU), Staudtstra{\ss}e 1, 91058 Erlangen, Germany, EU}
\affiliation{Max Planck Institute of Quantum Optics, Hans-Kopfermann-Stra{\ss}e 1, 85748 Garching, Germany, EU}
\author{Christoph Lemell}
\affiliation{Institute for Theoretical Physics, Vienna University of Technology, 1040 Vienna, Austria, EU}
\author{Georg Wachter}
\affiliation{Institute for Theoretical Physics, Vienna University of Technology, 1040 Vienna, Austria, EU}
\author{Florian Libisch}
\affiliation{Institute for Theoretical Physics, Vienna University of Technology, 1040 Vienna, Austria, EU}
\author{Thomas Madlener}
\affiliation{Institute for Theoretical Physics, Vienna University of Technology, 1040 Vienna, Austria, EU}
\author{Joachim Burgd\"orfer}
\affiliation{Institute for Theoretical Physics, Vienna University of Technology, 1040 Vienna, Austria, EU}
\author{Peter Hommelhoff}
\affiliation{Department of Physics, Friedrich-Alexander-Universit\"at Erlangen-N\"urnberg (FAU), Staudtstra{\ss}e 1, 91058 Erlangen, Germany, EU}
\affiliation{Max Planck Institute of Quantum Optics, Hans-Kopfermann-Stra{\ss}e 1, 85748 Garching, Germany, EU}
\affiliation{Max Planck Institute for the Science of Light, G\"unther-Scharowsky-Stra{\ss}e 1, Blg. 24, 91058 Erlangen, Germany, EU}

\begin{abstract}
We demonstrate coherent control of multiphoton and above-threshold photoemission from a single solid-state nanoemitter driven by a fundamental and a weak second harmonic laser pulse. Depending on the relative phase of the two pulses, electron emission is modulated with a contrast of the oscillating current signal of up to 94\%. Electron spectra reveal that all observed photon orders are affected simultaneously and similarly. We confirm that photoemission takes place within $10\,\mathrm{fs}$. Accompanying simulations indicate that the current modulation with its large contrast results from two interfering quantum pathways leading to electron emission.
\end{abstract}
\maketitle

Ionization by two-color laser fields with well-defined relative phase allows one to tune and control electronic dynamics on the (sub-) femtosecond time scale. By virtue of the synthesized laser field the energy and angular distributions of emitted electrons can be manipulated. Two-color pulses have been used in investigations of above-threshold ionization of atoms \cite{muller_above-threshold_1990,schumacher_phase_1994,ehlotzky_atomic_2001}, controlled ionization \cite{sheehy_phase_1995,thompson_one_1997,ohmura_coherent_2004}, dichroism \cite{fifirig_elliptic_2003,cionga_dichroic_2003} and orientation of molecules \cite{de_field-free_2009}. Recently, they have also been applied to control interference fringes in the momentum distribution of electron emission \cite{xie_attosecond_2012,arbo_interference_2014,arbo_effect_2014}. In this letter we demonstrate exquisite coherent emission control by simultaneous interaction of fundamental ($\omega$) and second harmonic ($2\omega$) femtosecond laser pulses with condensed matter. Specifically, we control photoemission from an individual nanotip.

While nanotips are nowadays routinely used as electron sources in high-resolution electron microscopes \cite{spence_high-resolution_2013}, their superb transverse coherence known from DC field emission has only recently been observed in photoemission \cite{ehberger_highly_2015}. Field enhancement at the apex of nanotips \cite{novotny_principles_2012} confines and enhances electron emission, thus enabling the study of strong-field physics with moderate laser intensities well below the damage threshold \cite{bormann_tip-enhanced_2010,schenk_strong-field_2010,kruger_attosecond_2011,kruger_interaction_2012,herink_field-driven_2012,piglosiewicz_carrier-envelope_2014}. Laser-induced photoemission from tips (for extensive reviews see \cite{hommelhoff_attosecond_2015}) has already been employed in pulsed electron guns for electron diffraction experiments \cite{gulde_ultrafast_2014,muller_femtosecond_2014}, while arrays of nanotips have lately been fabricated and explored as electron sources for compact coherent x-ray production \cite{graves_intense_2012,nagel_surface_2013,swanwick_nanostructured_2014}. Recently, a $\omega-2\omega$ experiment was performed on a silicon tip array, but no phase-resolved signal indicative of coherent control was reported \cite{swanwick_nanostructured_2014}.

In our experiment, atomic-scale in-situ control over the sample surface results in a well-defined nanoemitter that surpasses the limitations of focal averaging and inhomogeneous broadening. We show that electron emission induced by a strong fundamental pulse can be enhanced or suppressed with a contrast of up to 94\% when superimposing a weak second harmonic pulse. This scheme thereby allows efficient coherent control of the nanotip photocurrent utilizing the metal-vacuum interface to suppress emission in one direction. Accompanying simulations suggest that the strong modulation can be explained in terms of two nearly perfectly constructively or destructively interfering quantum pathways, each individually leading to electron emission.

\begin{figure}
\centering{\includegraphics[width=8cm]{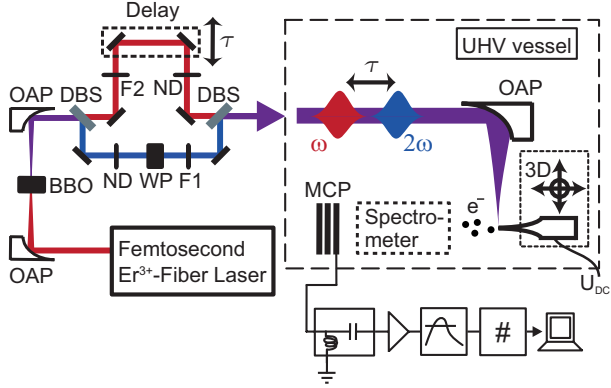}}
\caption{(Color online) Experimental setup. The second harmonic of Er$^{3+}$-doped fiber laser pulses is generated in BBO using off-axis parabolas (OAPs) for dispersion-free focusing and collimation. Beams enter a Mach-Zehnder interferometer with dichroic beam splitters (DBS), where the fundamental passes a variable delay stage. Intensities are independently controlled with neutral density wheels (ND) and the beams are spectrally filtered (F1, F2). The polarization of the second harmonic is rotated with a half-wave plate (WP). After recombination of the two pulses with variable time delay $\tau$ they are focused onto a tungsten tip with an off-axis parabola (OAP). Electrons are detected on a microchannel plate (MCP) or with a spectrometer using single electron pulse counting. 
\label{fig1}}
\end{figure}

\begin{figure*}
\centering{\includegraphics[width=0.75\textwidth]{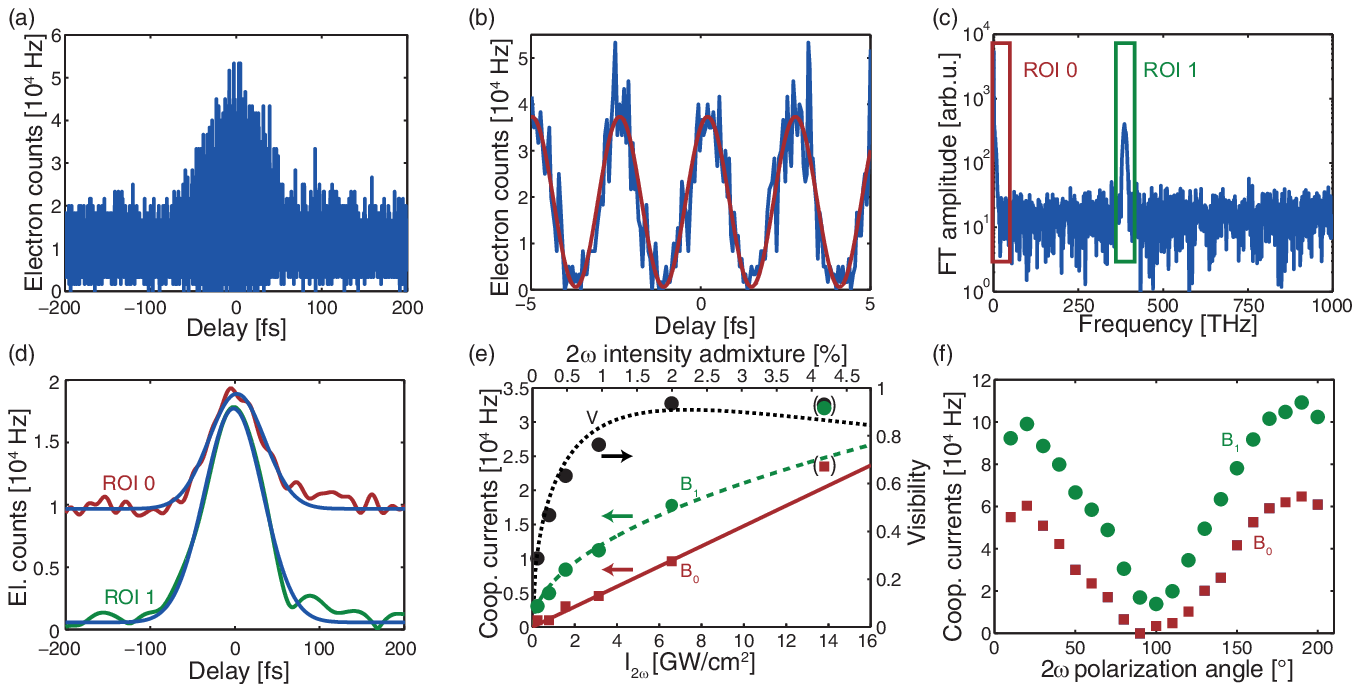}}
\caption{(Color online) Demonstration of coherent control of the photocurrent. (a) Electron current as a function of delay $\tau$ between $\omega$ and $2\omega$ pulse for $I_\omega = 330$ GW/cm$^2$ and $I_{2\omega} = 6.6$ GW/cm$^2$. For positive values of delay the $2\omega$ pulse encounters the tip first. We define $t=0$ at a current maximum in temporal overlap (b) Magnification of the central area of panel (a). A sine-fit to the data is shown as red solid line. (c) Absolute value of the Fourier transform (FT) of the data in (a). The cooperative photocurrent consists almost exclusively of an oscillatory component at $2\omega$ (ROI 1) and a DC component (ROI 0). (d) Inverse Fourier transformation (IFT) of data in ROI 0 and IFT and Hilbert transform of data in ROI 1. Gaussian fits to the data are shown in blue. (e) ROI amplitude fit parameters $B_0$ and $B_1$ as a function of the second harmonic intensity for a fundamental intensity of $I_\omega = 330$ GW/cm$^2$. $B_0$ and $B_1$ are proportional to $I_{2\omega}$ (red solid line) and $\sqrt{I_{2\omega}}$ (green dashed line), respectively. The contrast (visibility) of the current oscillation calculated using Eq.\,\ref{vis1} (black dots) and Eq.\,\ref{vis2} (black dotted line) reaches up to 94\% in this experiment. (f) $B_0$ (red squares) and $B_1$ (green spheres) as a function of the rotation angle $\theta$ of the polarization direction of the second harmonic with respect to the tip axis and the fundamental field for $I_\omega = 410$ GW/cm$^2$ and $I_{2\omega} = 13$ GW/cm$^2$. 
\label{fig2}}
\end{figure*}

We generate phase-locked pairs of fundamental and second harmonic pulses by passing the output of an Er$^{3+}$-doped femtosecond fiber laser through a 100 $\mu$m thick $\beta$-barium borate (BBO) crystal (Fig.\,\ref{fig1}). Central wavelengths and pulse durations are 1560 nm and 74 fs for the fundamental and 780 nm and 64 fs for the second harmonic. The pulses are sent into a Mach-Zehnder interferometer with dichroic beamsplitters where the delay (or, equivalently, the relative optical phase) between the pulses can be varied by up to 2 ps by a piezoelectric actuator and locked within this range \cite{wehner_scanning_1997}. The phase between the two colors is not measured in our experiment. A half-wave plate in the $2\omega$ arm rotates the $2\omega$ polarization to align it with the $\omega$ polarization and the tip axis, unless stated otherwise. Intensities can be adjusted independently by neutral density filter wheels. The laser pulses are directed into a UHV chamber and focused onto a (310)-oriented tungsten nanotip with a nominal work function of $W_0 = 4.31$ eV \cite{muller_work_1955}. With the applied bias voltage an effective barrier height of $W \approx 3.6$ eV results. The tips employed in the experiments display radii of curvature around 10\,nm, determined in situ by field ion microscopy \cite{muller_field_1965} (see SI for details \cite{supplementary_information}\nocite{klein1997,kruger_self-probing_2013,martin_wright_1960,zuber_an_2002,yudin_nonadiabatic_1965,vasp,PBE}). Photoelectrons from the biased tip are either counted directly at a microchannel plate (MCP) detector, or, alternatively, pass an energy high-pass filter and are only then detected by the MCP, enabling spectrally resolved measurements.

Typical peak near-field intensities at the tip's apex are on the order of $I_\omega=4\cdot 10^{11}\,\mathrm{W/cm}^2$ for the fundamental and $I_{2\omega}=6\cdot 10^9\,\mathrm{W/cm}^2$ for the second harmonic, including field enhancement \cite{thomas_probing_2013,thomas_large_2015} (see SI for details). The near-fields of both colors result in electron emission in the multi-photon regime with minimum Keldysh parameters $\gamma = \sqrt{W_0/2U_p}$ of $\gamma_{\omega,\mathrm{min}} = 2.7$ and $\gamma_{2\omega,\mathrm{min}} = 47$, where $U_p$ is the ponderomotive energy. The second harmonic is much weaker than the fundamental with the ratio of second harmonic to fundamental near-field peak intensity $I_{2\omega}/I_\omega\leq 4$\%. Electron emission induced exclusively by a pulse at frequency $\omega$ is observed to be at least more than one order of magnitude larger than electron emission by a $2\omega$-pulse.

In regions of temporal overlap of the $\omega$ and $2\omega$ pulses we observe a dramatic change in the emission characteristics: The emitted current strongly oscillates as a function of the delay (central region of Fig.\,\ref{fig2}(a), magnified in panel (b)). This behavior is well described by a sine fit (red solid line). The contrast is up to 94\%, i.e., electron emission is either drastically enhanced or reduced to almost zero as compared to the emission from the tip for separated pulses. Despite the weak $2\omega$ intensity admixture of $2\,\%$ the maximum achievable electron current is almost four times the current for separated pulses for these parameters.

For further analysis we Fourier transform the data of Fig.\,\ref{fig2}(a) and obtain the spectrum shown in Fig.\,\ref{fig2}(c). On top of a white-noise background the Fourier spectra for overlapping pulses feature two main peaks: a low-frequency (DC) component in region of interest 0 (ROI 0) and a component at finite frequency (ROI 1) peaking at 390 THz, the frequency of the second harmonic. A windowed inverse Fourier transform of the data in ROI 0 and ROI 1 and an additional Hilbert transform of the data in ROI 1 to obtain its envelope yield the temporal variation of the amplitude of the ROI 0 and ROI 1 components as a function of the delay $\tau$ (Fig.\,\ref{fig2}(d)). These can be well approximated by Gaussian fits $G_i(\tau) = A_i+B_i\exp (-4 \ln{2}\tau^2/\Delta_i^2)$. Here $i=0,1$ for ROI 0 and ROI 1, $\tau$ denotes the $\omega-2\omega$ delay, and $\Delta_i$ are the full widths at half maximum of the photocurrent components. The ROI amplitude fit parameters $B_i$ serve as direct measure of the cooperative effects in the region of temporal overlap.

We find that $B_0$ and $B_1$ scale differently with the second harmonic intensity $I_{2\omega}$. While $B_0$ increases linearly with $I_{2\omega}$, $B_1$ shows a $\sqrt{I_{2\omega}}$ dependence (Fig.\,\ref{fig2}(e)). The contrast of the oscillating current signal, i.e. its visibility
\begin{equation}
V = \frac{N_{\max} - N_{\min}}{N_{\max} + N_{\min}}\label{vis1}
\end{equation}
determined from sine fits reaches up to 94\% for an intensity admixture of the $2\omega$-component of 2\% to the fundamental near-field intensity of $I_\omega=330$ GW/cm$^2$ (Fig.\,\ref{fig2}(e)). With the emergence of higher-order contributions (additional Fourier component at $4\omega$) for admixtures of $4\%$ the second harmonic cannot be regarded a weak perturbation any longer. We therefore exclude the data point at $I_{2\omega}=14$ GW/cm$^2$ from the following analysis.

The cooperative electron emission also varies as a function of the orientation of the incident polarization vector of the second harmonic pulse (Fig.\,\ref{fig2}(f)). The maximum cooperative effect is found for the $2\omega$ polarization aligned with the $\omega$ component, which is parallel with the tip axis. Rotating the incident $2\omega$ polarization vector the cooperative electron emission is reduced and reaches its minimum for the $2\omega$ polarization perpendicular to the tip axis. This is expected as the near-field excited at the apex by the $2\omega$ field is strongest for an incident polarization aligned with the tip axis and also overlaps best with the near-field excited by the fundamental. Further, for an incident polarization vector orthogonal to the tip axis its enhanced region is moved away from the tip apex \cite{yanagisawa_optical_2009}, so away from the low-workfunction (310) plane.

\begin{figure}
\centering{\includegraphics[width=8cm]{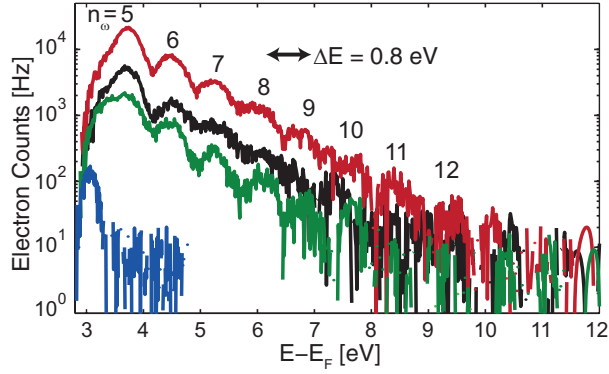}}
\caption{(Color online) Experimental electron spectra for $I_\omega = 410$ GW/cm$^2$ and $I_{2\omega} = 13$ GW/cm$^2$. Blue: $2\omega$ pulse alone. Black: $\omega$ pulse alone. Red: $\omega$ and $2\omega$ pulses, with relative phase locked at photocurrent maximum. Green: $\omega$ and $2\omega$ pulses with relative phase locked at photocurrent minimum.}
\label{fig3}
\end{figure}

To gain further insight into the underlying processes we have recorded electron-energy spectra when either one of the two pulses is blocked or both are simultaneously present, see Fig.\,\ref{fig3}.
With only the $2\omega$ pulse on target (blue line) we observe only a weak two-photon emission current. By contrast, with only the fundamental pulse present (black line) a typical above-threshold photoemission (ATP) spectrum containing many orders appears \cite{schenk_strong-field_2010}. Overlapping $\omega$ and $2\omega$ pulses can homogeneously suppress or enhance emission for the pulse delay locked to minimum (green) and maximum (red) of total photocurrent. For quantitative analysis of the two-color emission we decompose the total emission rate $R_{tot}=\sum_n{r_n}$ and record the individual emission rates $r_n$ of several multiphoton orders $n$ as a function of two-color phase $\phi$. We fit them with sinusoidal fits of the form $r_n(\phi)=C_n+D_n \cos(\phi+\phi_n)$. We thus obtain individual visibilities $V_n=D_n/C_n$ and phase offsets $\phi_n$. Our results show that the emission rates of all multiphoton orders vary sinusoidally with the two-color phase, have the same visibility corresponding to the overall photocurrent visibility, and display no relative phase offset (see section II of the SI for details).

Therefore, the total emitted current can be described by just two effective emission pathways (see section III of SI). The emitted current from fundamental alone ($\approx A_0$) scales with $I_\omega^4$ despite the presence of higher photon orders. Thus, effective emission pathway 1 with rate $R_1$ corresponds to the absorption of four photons of the fundamental. Effective pathway 2 involves the absorption of two photons of the fundamental and one photon of the second harmonic resulting in electron emission with rate $R_2$. Thereby the second harmonic is treated as a (lowest-order) perturbation. The interference term between the two paths $R_{12}$ oscillates with the relative phase between the pathways, $\phi(\tau) = \phi_{2\omega} - 2\phi_\omega - \phi_e$, which we control via the adjustable pulse delay $\tau$. $\phi_e$ denotes the unknown phase relation of the electronic states connected by the optical transition. It is noteworthy that our model is reminiscent of heterodyning. It is easy to see that this model predicts the following dependencies on the intensities:
\begin{eqnarray}
R_1&=&\alpha^4 I_\omega^4\\
R_2&=&\alpha^2 I_\omega^2 \cdot \beta I_{2\omega} \\
R_{12}&=&2\sqrt{R_1R_2} \cos\phi = 2\alpha^3 I_\omega^3 \sqrt{\beta I_{2\omega}} \cos\phi,
\end{eqnarray}
where $\alpha$ and $\beta$ are proportionality factors for the absorption of $\omega$ and $2\omega$ photons, respectively. We identify the cooperative contribution to $B_0$ with pathway $R_2$ and the amplitude of the oscillation $B_1$ with $R_{12}$.

The parameters $\alpha$ and $\beta$ can be extracted by comparing our experimental data with the quantum-pathway interference model introduced above. Evaluating $R_1$ ($\propto\alpha^4$) from the observed count rate with the fundamental at $I_\omega = 330$ GW/cm$^2$ alone we find $\alpha^4 = (8.068 \pm 0.080)\cdot 10^{-43}$ Hz cm$^8$ W$^{-4}$. Evaluating $R_2$ to extract $\alpha^2 \beta$ from the scaling of $B_0$ with second harmonic intensity in Fig.\,\ref{fig2}(e) yields $\alpha^2 \beta = (1.22 \pm 0.15)\cdot 10^{-29}$ Hz cm$^6$ W$^{-3}$. From these results for $\alpha^4$ and $\alpha^2 \beta$ we can directly obtain the prefactor for the interference term $R_{12}$, yielding $2 \alpha^3\sqrt{\beta} = (6.27 \pm 0.38)\cdot 10^{-36}$ Hz cm$^7$ W$^{-3.5}$. This factor can also be derived independently from the scaling of $B_1$ with $I_{2\omega}$ (Fig.\,\ref{fig2}(e)), which yields $2 \alpha^3\sqrt{\beta} = (5.70 \pm 0.52)\cdot 10^{-36}$ Hz cm$^7$ W$^{-3.5}$. The excellent agreement between the two independently obtained values for $2\alpha^3 \sqrt{\beta}$ is a strong indication that our simple model captures the essential features of the two-color ionization process well.

We can furthermore evaluate the visibility of the interference fringes,
\begin{equation}
V(I_\omega, I_{2\omega})=\frac{R_{12}}{R_1+R_2}=\frac{2\alpha^3 I_\omega^3 \sqrt{\beta I_{2\omega}}}{\alpha^4 I_\omega^4+\alpha^2 I_\omega^2 \beta I_{2\omega}}\, .\label{vis2}
\end{equation}
Fig.\,\ref{fig2}(e) shows the visibility $V(I_\omega, I_{2\omega})$ calculated using the quantum-pathway interference model (black dotted line) together with the experimental data points (Eq.\,\ref{vis1}). Again, excellent agreement between the model and the experimental data is found.

\begin{figure}
\centering{\includegraphics[width=8cm]{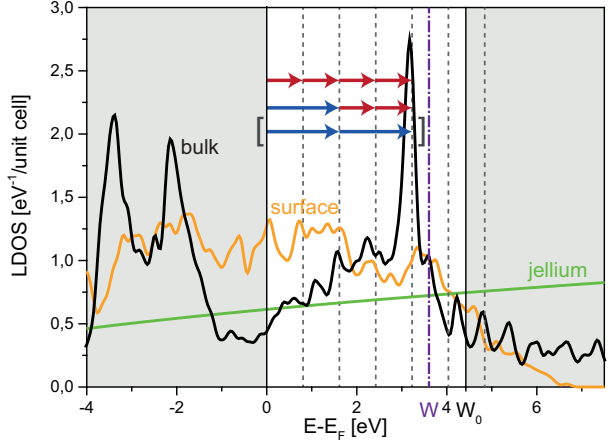}}
\caption{(Color online) Local density of states from a ground-state slab calculation (20 layers) for W(310). LDOS at surface (labelled surface, orange), LDOS for jellium (labelled jellium, green), and for comparison, DOS from a bulk calculation (labelled bulk, black). Intermediate surface and volume states are found at $2\hbar\omega=1.6$ eV and $4\hbar\omega=3.2$ eV. Three pathways to an energy of $E_F+4\hbar\omega$ are indicated, where the lowest is negligible under our conditions because of the small $2\omega$ intensity. The reduction of the workfunction due to the bias voltage is indicated by the purple dash-dotted line at 3.6\,eV.
\label{fig4}}
\end{figure}

Both a simple tunneling model and a 1D jellium time-dependent density functional theory (TDDFT) simulation \cite{kruger_interaction_2012} predict a strong current modulation with phase delay but fail to describe the sinusoidal shape observed in the experiment (see section V of SI for details). In a 3D ground-state density functional theory (DFT) calculation for the W(310) surface we observe a strong modulation of the local density of states (LDOS) below the vacuum level compared to the smooth $\rho\propto\sqrt{E}$ of the jellium model (Fig.\ \ref{fig4}, see SI for details).
Of importance appear the pronounced peaks in the vicinity of $E_F+2\hbar\omega$ and in particular $E_F+4\hbar\omega$. These broad peaks can act as doorway states resonantly enhancing multiphoton emission compared to the prediction of the jellium model. Their presence supports the quantum-pathway interference model for reaching the short-lived resonance at $E_F+4\hbar\omega$, from where photoemission proceeds.

The lifetime of excited states is limited by their spectral width, by the presence of the optical field, which ionizes them, and by diffusion of the electron to the bulk. From our experiment we can determine an upper bound for the lifetime of involved intermediate states to be less than 10\,fs. This can be concluded from the full width at half maximum of the photocurrent components, which are $(78 \pm 10)$\,fs for ROI 0 and $(84 \pm 10)$\,fs for ROI 1, determined from the Gaussian fits to the data in Fig.\,\ref{fig2}. These are slightly shorter than the expected cooperative signals from a simple convolution of the laser pulses for all applied intensities. The symmetry of the photocurrent components of ROI 0 and ROI 1 as a function of the pulse delay is another indicator of a short lifetime.

In summary, we have demonstrated quantum-pathway interference in multiphoton and above-threshold photoemission at a tungsten tip on femtosecond time scales with two-color femtosecond driving fields. Despite the solid-state nature of the emitter, the maximum modulation depth of 94\% is, to the best of our knowledge, the highest reported value for two-color photoemission experiments, including those for isolated atoms in the gas phase \cite{ackermann_strong_2014}. This high visibility is closely related to the nanometer-sized interaction volume, preventing the influence of focal averaging effects and inhomogeneous broadening. Nanotips may therefore be used as nanometric probes to measure light phases \cite{chen_measurements_1990}. Moreover, we demonstrate a potential building block for the development of lightwave electronics \cite{krausz_attosecond_2014} on the femtosecond duration and nanometer length scales. Additionally, we expect that polarization-shaped \cite{brixner_femtosecond_2001} two-color laser pulses together with the polarization-sensitive near-field distribution \cite{yanagisawa_optical_2009}, the spatially-dependent work function on the nanometer scale, and the Gouy phase shift will permit engineered spatiotemporal electron emission profiles at single tips and tip arrays, highly desired for use in electron guns for coherent x-ray generation \cite{graves_intense_2012}.

\begin{acknowledgments}
This project was funded in part by the ERC grant ``Near Field Atto'', by DFG SPP 1840 ``QUTIF'', and by the Austrian Science Fund (FWF) within the special research projects SFB-041 ``ViCoM'' and SFB-049 ``Next Lite'' and project P21141-N16. M.F. and G.W. acknowledge support by the IMPRS-APS. We thank Sebastian Thomas for supporting FDTD simulations, Alexander Lang for contributions to the experimental setup, and Misha Yu. Ivanov, Thomas Fauster, Martin Hundhausen, and Takuya Higuchi for helpful discussions.
\end{acknowledgments}

\bibliography{two-color-bibliography}
\end{document}